\documentclass[prd,review,nofootinbib,notitlepage,aps,tightenlines,amsmath,amssymb,showpacs,superscriptaddress]{revtex4-1}

\usepackage{amsmath}
\usepackage{amssymb}
\usepackage{dcolumn}
\usepackage{bm}
\usepackage{feynmp}                                     
\usepackage{hyperref,graphicx}           
\usepackage[utf8]{inputenc}
\usepackage{caption}
\usepackage{subcaption}
\usepackage{color}
\def\beq{\begin{equation}}
\def\eeq{\end{equation}}
\def\bea{\begin{eqnarray}}
\def\eea{\end{eqnarray}}
\def\no{\nonumber}
\def\nn{\nonumber\\}

\begin{document}

\title{Detecting a Secondary Cosmic Neutrino Background from Majoron 
Decays in Neutrino Capture Experiments}

\author{Zackaria Chacko}
\author{Peizhi Du}
\affiliation{Maryland Center for Fundamental Physics, Department of Physics,\\ University of Maryland, College Park, MD 20742-4111 USA}
\author{Michael Geller}
\affiliation{Maryland Center for Fundamental Physics, Department of Physics,\\ University of Maryland, College Park, MD 20742-4111 USA}
\affiliation{Department of Physics, Tel Aviv University, Tel Aviv 6997801, Israel}

\date{\today}
\begin{abstract}

 We consider theories in which the generation of neutrino masses is 
associated with the breaking of an approximate global lepton number 
symmetry. In such a scenario the spectrum of light states includes the 
Majoron, the pseudo-Nambu Goldstone boson associated with the breaking 
of the global symmetry. For a broad range of parameters, the Majoron 
decays to neutrinos at late times, after the cosmic neutrinos have 
decoupled from the thermal bath, resulting in a secondary contribution 
to the cosmic neutrino background. We determine the current bounds on 
this scenario, and explore the possibility of directly detecting this 
secondary cosmic neutrino background in experiments based on neutrino 
capture on nuclei. For Majoron masses in the eV range or below, the 
neutrino flux from these decays can be comparable to that from the 
primary cosmic neutrino background, making it a promising target for 
direct detection experiments. The neutrinos from Majoron decay are 
redshifted by the cosmic expansion, and exhibit a characteristic energy 
spectrum that depends on both the Majoron mass and its lifetime. For 
Majoron lifetimes of order the age of the universe or larger, there is 
also a monochromatic contribution to the neutrino flux from Majoron 
decays in the Milky Way that can be comparable to the diffuse 
extragalactic flux. We find that for Majoron masses in the eV range, 
direct detection experiments based on neutrino capture on tritium, such 
as PTOLEMY, will be sensitive to this scenario with 100 gram-years of 
data. In the event of a signal, the galactic and extragalactic 
components can be distinguished on the basis of their distinct energy 
distributions, and also by using directional information obtained by 
polarizing the target nuclei.

\end{abstract}

\preprint{}



\maketitle

\section{Introduction\label{s.intro}}

 While oscillation experiments have conclusively established that 
neutrinos have small but non-vanishing masses, the dynamics that 
generates these masses remains a mystery. In particular, it is not known 
whether neutrinos are Dirac or Majorana particles, whether their 
spectrum is hierarchical, inverse hierarchical or quasi-degenerate, or 
why their masses are so small. Several ideas have been put forward to 
explain the origin of neutrino masses. These include the well-known 
seesaw mechanism 
\cite{Minkowski:1977sc,Yanagida,Glashow,GellMann:1980vs,Mohapatra:1980yp}, 
and the Majoron model 
\cite{{Gelmini:1980re},{Chikashige:1980ui},{Georgi:1981pg}}.

Extensions of the SM based on the Majoron framework incorporate a lepton 
number as a global symmetry. Neutrinos acquire masses once the global 
lepton number symmetry is spontaneously broken. The Majoron is the 
pseudo-Nambu Goldstone boson associated with the breaking of the global 
symmetry. The form of its couplings to neutrinos is dictated by the 
non-linearly realized lepton number symmetry. In the case when neutrinos 
are Dirac, the couplings of the Majoron field $J$ take the schematic 
form,
 \begin{equation}
i \frac{m_\nu}{f} J \nu n^c + {\rm h.c.}
\label{DiracCoupling}
 \end{equation}
 Here $m_\nu$ represents the neutrino mass, while $f$ denotes the 
Majoron decay constant, which corresponds to the scale at which the 
global lepton number symmetry is broken. In the case when the neutrino 
masses are Majorona, the corresponding coupling takes the form
 \begin{equation}
i \frac{m_\nu}{f}\frac{ J}{2} \nu \nu + {\rm h.c.}
\label{MajoranaCoupling}
 \end{equation} 
 Eqs.~(\ref{DiracCoupling}) and (\ref{MajoranaCoupling}) represent the 
leading order interactions of the Majoron with neutrinos in an expansion 
in $1/f$. These couplings are diagonal in the basis in which the 
neutrino masses are diagonal. Hence the Majoron couples to each neutrino 
mass eigenstate with a strength proportional to the corresponding 
neutrino mass.

In general, the global symmetry need not be exact but only approximate, 
in which case the Majoron will acquire a mass. It then follows from the 
couplings of the Majoron to neutrinos, Eqs.~(\ref{DiracCoupling}) and 
(\ref{MajoranaCoupling}), that any population of Majorons present in the 
early universe will eventually decay to neutrinos, unless such decays 
are forbidden by phase space considerations. If the Majoron lifetime is 
greater than the age of the universe, Majorons could constitute some or 
all of the observed dark matter of the universe \cite{Rothstein:1992rh, 
Berezinsky:1993fm, Frigerio:2011in, Rojas:2017sih, Brune:2018sab}. The 
required population of Majorons could have survived till the present day 
as thermal relics, or have been produced nonthermally through, for 
example, the misalignment mechanism. If, however, the Majoron lifetime 
is less than or comparable to the age of the universe, decays of the 
Majoron will result in a new contribution to the population of cosmic 
neutrinos. Provided the Majoron decays occur at late times, after the 
cosmic neutrinos have decoupled from the thermal bath, and at 
temperatures below the mass of the Majoron, this new population of 
neutrinos will be thermally decoupled and distinct from the cosmic 
neutrino background (C$\nu$B). Several authors have explored the 
possibility of detecting the neutrinos from Majoron decays in neutrino 
detectors such as Borexino, Super-Kamiokande and IceCube 
\cite{PalomaresRuiz:2007ry, Garcia-Cely:2017oco, Covi:2009xn}, as well 
as in dark matter detectors (see, for example, \cite{Cui:2017ytb}). In 
this paper we determine the current bounds on this scenario, and explore 
the possibility of directly detecting this secondary C$\nu$B in neutrino 
capture experiments.

Cosmological observations can be used to place limits on the flux of 
neutrinos from Majoron decay. These limits depend on the Majoron mass 
and lifetime. Decays that occur prior to recombination result in a 
contribution to the energy density in radiation that is present during 
the epoch of acoustic oscillations. Precision measurements of the the 
cosmic microwave background (CMB) place severe limits on any such 
contribution, with the result that the number density of neutrinos from 
Majoron decays is always expected to be smaller than the C$\nu$B. In the 
case of Majorons that decay after recombination, the limits are much 
weaker. Current cosmological observations allow roughly 5\% of the total 
energy density in dark matter to have decayed to dark radiation before 
today \cite{Poulin:2016nat, Chudaykin:2016yfk, 
Chudaykin:2017ptd}.\footnote{The bound on the fraction of decaying dark 
matter varies with the lifetime. The current bound indicates $ 
\tau/\mathcal{F}\gtrsim 160 \,\textrm {Gyr}$ for $\tau\gtrsim t_0$ 
\cite{Poulin:2016nat}, where $\tau$ is the lifetime of the decaying dark 
matter while $\mathcal{F}$ denotes the fractional of energy density in 
decaying dark matter to that of all dark matter. Since we focus on the 
case in which the lifetime is order of the age of universe, we take the 
bound to be $\sim5\%$. } Then, if neutrinos are Majorana, the number 
density of neutrinos and antineutrinos from Majoron decay is given by
 \begin{equation}
 n^J_{\nu} = n^J_{\bar{\nu}} = \mathcal{F} \Omega_{\rm dm} \frac{\rho_c}{m_J}
\left( 1 - e^{-{t_0}/{\tau_J}}\right) \;.
 \end{equation}
 In this expression $\tau_J$ represents the lifetime of the Majoron, and 
$m_J$ its mass. $\mathcal{F}$ denotes the fractional contribution of 
Majorons to the total energy density in dark matter at early times, 
while $\Omega_{\rm dm}$ is the fractional contribution of dark matter to 
the total energy density $\rho_c$. If the Majoron lifetime is much less 
than $t_0$, the age of the universe, we have
 \begin{equation}
 n^J_{\nu0} = n^J_{\bar{\nu}0} = 
\mathcal{F} \Omega_{\rm dm} \frac{\rho_c}{m_J} = 
\frac{63}{{\rm cm}^3} \frac{\mathcal{F}}{5\%} \frac{{\rm eV}}{m_J}  \;.
 \end{equation}
 If instead neutrinos are Dirac, the number of active neutrinos and 
antineutrinos is half of this. For comparison, note that the total 
number density of the three flavors of neutrinos in the C$\nu$B is 
$168/\textrm{cm}^3$, with an equal number of antineutrinos. We see from 
this that the number density of neutrinos arising from Majoron decays 
can be very large. In particular, for Majoron masses below an eV, the 
flux from their decays can be comparable to or even exceed the flux from 
the cosmic background neutrinos. Therefore, the neutrinos from Majoron 
decays constitute a promising target for the direct detection 
experiments designed to detect the C$\nu$B.

Since there are only two particles in the final state, the neutrinos 
produced in Majoron decays are monochromatic. However, as a consequence 
of the cosmic expansion, these particles get redshifted, resulting in a 
characteristic energy distribution that depends not just on the Majoron 
mass, but also its width. If the Majoron lifetime is of order the age of 
the universe, we also expect a flux of neutrinos from the decay of 
Majorons in the Milky Way galaxy. Since the time taken to traverse the 
galaxy is small compared to the age of the universe, these neutrinos are 
monochromatic. Provided the population of Majorons is highly 
nonrelativistic during the matter dominated era, the distribution of 
Majorons in the galaxy is expected to follow the cold dark matter 
profile. Then the overdensity of Majorons in the Milky Way results in a 
flux of galactic neutrinos that is comparable to the diffuse 
extragalactic flux.

 Experiments to detect the C$\nu$B based on neutrino capture on tritium 
were first proposed by Weinberg more than 50 years ago 
\cite{Weinberg:1962zza}. Tritium undergoes beta decay with a half-life 
of 12.32 years,
 \begin{equation}
^3{\rm H} \rightarrow \; ^3{\rm He} + e^- + \bar{\nu}_e   \; \; .
 \end{equation}
 This process releases 18.6 keV of energy, which is distributed between 
the final state electron and neutrino. However, this decay can be also 
induced by neutrino capture,
 \begin{equation}
^3{\rm H} + \nu_e \rightarrow \; ^3{\rm He} + e^-    \; \; .
 \end{equation}
 The final state electron now carries away the energy of the incident 
neutrino, in addition to the energy liberated in the decay. Therefore, 
given sufficient energy resolution, searches for electrons with energies 
above the beta decay endpoint can be used for direct detection of cosmic 
neutrinos.

Recently, this idea has taken concrete shape in the form of the PTOLEMY 
experiment \cite{Betts:2013uya, Baracchini:2018wwj}, which aims to 
discover the C$\nu$B. PTOLEMY proposes to use 100g of tritium coated on 
graphene. A target of this size could detect about 4 C$\nu$B events per 
year if neutrinos are Dirac, rising to 8 events per year if neutrinos 
are Majorana \cite{Long:2014zva}. The energy resolution $\Delta$ is at 
the level of 0.15 eV \cite{Betts:2013uya}, which means that the 
experiment is not sensitive to neutrinos with energies below this 
threshold. The C$\nu$B neutrinos are non-relativistic. Therefore, in 
order to detect them, the neutrinos must be quasi-degenerate, with 
masses of order 0.1 eV.

In the case of Majorons with masses below an eV, the flux of neutrinos 
from Majoron decay can be comparable to or larger than the flux from the 
C$\nu$B.\footnote{For models of new physics that alter the number 
density of nonrelativistic cosmic neutrinos, see 
for example \cite{Chen:2015dka, Zhang:2015wua, Arteaga:2017zxg}.} We 
find that for Majoron masses in the eV range, direct detection 
experiments based on neutrino capture on tritium will be sensitive to 
this scenario with 100 gram-years of data even if the SM neutrinos are 
lighter than $0.1\,$eV. Apart from an effect on the total event rate, we 
expect a distinctive energy spectrum of events inherited from the 
characteristic energy distributions of the neutrinos from extragalactic 
and galactic decays. These signals can be distinguished from the 
C$\nu$B, because the C$\nu$B neutrinos are all nonrelativistic today.

Interestingly, by employing polarized tritium nuclei as the target 
\cite{Lisanti:2014pqa}, we can also obtain directional information about 
the neutrino flux. Since galactic Majorons are expected to follow the 
dark matter distribution in the Milky Way, the monochromatic component 
of the neutrino flux encodes information about the profile of dark 
matter in the galaxy. Therefore, by observing the modulation of the 
event rate with the orientation of the tritium spin, we can shed light 
on the galactic dark matter distribution.

This paper is organized as follows. We introduce the Majoron model in 
Sec.~\ref{s.majoron}, and discuss the current bounds on this class of 
theories. In Sec.~\ref{s.flux} we study the neutrino flux from Majoron 
decay, including both extragalactic and galactic sources. In 
Sec.~\ref{s.signal}, we estimate the signal rates at PTOLEMY based on 
two benchmark points. We also study the dependence of the signal on the 
orientation of the spin of the tritium nuclei in the case of a polarized 
target. We conclude in Sec.~\ref{s.conclusion}.

\section{The Majoron Model\label{s.majoron}}

As the Goldstone boson associated with the breaking of lepton number, 
the form of the Majoron couplings in the low energy theory is dictated 
by the non-linearly realized symmetry. If neutrinos are Dirac, below the 
symmetry breaking scale $f$, the coupling of the Majoron to leptons 
takes the schematic form,
 \begin{equation}
i\lambda \frac{J}{f} HLn^c + {\rm h.c.} \;
\label{uvdirac}
  \end{equation}
 In this equation the dimensionless parameter $\lambda$ is related to 
the neutrino mass as $\lambda = \sqrt{2} m_\nu/v_{\rm EW}$, where 
$v_{\rm EW} = 246$ GeV is the electroweak VEV. In the case when 
neutrinos are Majorana, we have instead
 \begin{equation}
i\frac{J}{f} \frac{(HL)^2}{\Lambda} + {\rm h.c.} \;
\label{uvmajorana}
 \end{equation}
 Here $\Lambda$ is an ultraviolet scale that is related to the neutrino 
mass as $\Lambda^{-1} = m_\nu/v_{\rm EW}^2$. Eqs.~(\ref{uvdirac}) and 
(\ref{uvmajorana}) are the leading order interactions of the Majoron 
with neutrinos in an expansion in $1/f$. At low energies these reduce to 
Eq.~(\ref{DiracCoupling}) and Eq.~(\ref{MajoranaCoupling}) respectively.

Assuming $m_J \gg m_\nu$, we obtain the total decay width of the Majoron to 
neutrinos as
 \bea
\Gamma^{D}_{J}=\frac{m_\nu^2}{f^2}\frac{m_J}{8\pi}~~~\textrm{or }~~\Gamma^{M}_{J}=\frac{m_\nu^2}{f^2}\frac{m_J}{16\pi} \;,
 \eea
 where $\Gamma^D_J $ and $\Gamma^M_J$ correspond to the Dirac and 
Majorana cases respectively. We focus on Majoron masses $m_J$ of order 
an eV and Majoron lifetimes $\tau_J$ of the order of the age of the 
Universe. This translates to a scale $f$ of order $10^6$ GeV.

The Majoron model is constrained by astrophysical, cosmological and 
collider data. These bounds can be translated into limits on the mass of 
the Majoron $m_J$ and the decay constant $f$. For light Majorons, with 
masses less than an MeV, the bounds from cosmology are the most severe. 
We therefore begin by considering the cosmological limits. Precision 
measurements of the CMB place limits on the total energy density in 
radiation during the epoch of acoustic oscillations. Additional energy 
density in radiation beyond the SM prediction is generally parametrized 
in terms of the effective number of neutrinos, $\Delta N_{\rm eff}$. The 
CMB constraint, $\Delta N_{\rm eff} < 0.3$ \cite{Aghanim:2018eyx}, 
translates into the requirement that the Majoron not have a thermal 
abundance at temperatures of order an MeV, when the weak interactions 
decouple. For low values of $f$, Majoron-neutrino scattering will bring 
the Majoron into thermal equilibrium with the neutrinos. To satisfy the 
bound, the decay constant $f$ must lie above 100 MeV for $m_J\lesssim 
1\,$MeV in the case of Majorana neutrinos~\cite{Chacko:2003dt}. The 
limits in the Dirac case, though somewhat model dependent, are 
comparable.

For Majorons with masses around eV, an even more severe constraint 
arises from the requirement that the Majoron not be in thermal 
equilibrium with the neutrinos during the CMB epoch. Inverse decays into 
Majorons prevent the neutrinos from free streaming, impacting the 
heights and locations of the CMB peaks. Current bounds indicate that 
neutrinos must be free streaming at temperatures $T$ of order an 
eV~\cite{Archidiacono:2013dua, Forastieri:2015paa}. This translates into 
a constraint on the Majoron decay constant, $f \gtrsim 100$ GeV 
\cite{Chacko:2003dt}, in the case of Majorana neutrinos. Again, the 
limits in the Dirac case, though somewhat model dependent, are similar.

So far we have focused on the cosmological limits arising from the 
dominant decay channel of Majorons, the decay to two neutrinos. Other 
subdominant channels, such as decays charged leptons and photons, can 
also be used to set limits on $f$ because the bounds on fluxes of 
visible particles are usually stronger. Since our focus is on Majoron 
masses in the eV range, the only relevant channel is $J\to 
\gamma\gamma$. However, for light Majorons the branching ratio to 
photons is extremely small, and the resulting bounds are very 
weak~\cite{Bazzocchi:2008fh, Lattanzi:2013uza, Garcia-Cely:2017oco}.

The strongest astrophysical bounds arise from the effects of Majorons on 
supernovae. Neutrinos in a supernova acquire masses from matter effects, 
which are largest for electron neutrinos. Then, if the Majoron is light, 
electron neutrinos can decay into lighter flavors of neutrinos and 
Majorons. This can affect a supernova in two distinct ways. Prior to the 
bounce, these decays can deleptonize the core, preventing the explosion 
from taking place. After the bounce, the Majorons produced in these 
decays can free stream out, leading to overly rapid energy loss. While 
the resulting constraints depend on the details of supernova dynamics, 
they are at the level of $f \gtrsim 100$ keV~\cite{Gelmini:1982rr, 
Kolb:1987qy, Choi:1989hi, Kachelriess:2000qc, Farzan:2002wx}, and 
therefore weaker than the cosmological bounds. Clearly, our benchmark 
values of $f\sim 10^6$ GeV and $m_J\sim 1$ eV are safe from all current 
bounds.

\section{The Neutrino Flux from Majoron Decays\label{s.flux}}

 In this section we determine the flux and energy distribution of the 
neutrinos arising from Majoron decays, as a function of the initial 
abundance of the Majoron, its mass and lifetime. In the case of Majorons 
with lifetimes longer than or comparable to the current age of the 
universe we expect two distinct contributions to the flux, one from 
Majoron decays in the Milky Way galaxy, and the other from extragalactic 
decays. The galactic flux is expected to be monochromatic, with 
frequency equal to exactly half the Majoron mass. The diffuse 
extragalactic component of the flux, however, undergoes redshift as the 
universe expands, resulting in an energy distribution that, although not 
monochromatic, is highly distinctive.

 \subsection{Extragalactic Neutrinos }\label{sec:e.g._neutrino}
 We first determine the number density of diffuse neutrinos arising from 
Majoron decaying outside the Milky Way. Assuming SM neutrinos are 
Majorana, the comoving number density of neutrinos and antineutrinos 
arising from Majoron decay in a time interval $dt$ is given by,
 \begin{equation}
 d(n_\nu a^3) = d(n_{\bar\nu} a^3)= -\frac{n_J a^3}{\tau_J}  dt \; ,   
\label{nurate}
 \end{equation}
 where $a$ represents the scale factor and $n_J a^3$ the comoving
number density of Majorons. Now $n_J$ evolves as
 \begin{equation}
n_J = n_{J0} e^{-{t}/{\tau_J}} \; ,
 \end{equation}
 where $n_{J0}$ is the initial number density of Majorons. If Majorons 
contribute a fraction $\mathcal{F}$ of the energy density in dark matter 
at early times, we have
 \begin{equation}
n_{J0} a^3 = \mathcal{F} \Omega_{\rm dm} \frac{\rho_c}{m_J}
 \end{equation}
 Then, Eq.~(\ref{nurate}) becomes
 \begin{equation}
 d(n_\nu a^3) = -\frac{\mathcal{F} \Omega_{\rm dm}}{\tau_J} \frac{\rho_c}{m_J} e^{-{t}/{\tau_J}} dt
\label{nu0}
 \end{equation}
 At times when the universe dominated by dark matter and dark energy, 
the Hubble expansion at redshift $z$ can be approximated as
 \begin{equation}
 H(z) = H_0 \sqrt{\Omega_\Lambda + (1 + z)^3 \Omega_m} \; .
\label{Hz}
 \end{equation}
 Here $\Omega_\Lambda$ and $\Omega_m$ denote the fractional 
contributions of dark energy and matter to the total energy density of 
the universe, and we are working in the limit $\Omega_\Lambda + \Omega_m 
= 1$. We can use Eq.~(\ref{Hz}) to obtain an expression for the age of 
the universe at redshift $z$,
 \begin{equation}
t(z) = \frac{2}{3} \frac{1}{H_0 \sqrt{\Omega_\Lambda}}
{\rm ln}\left(\sqrt{r(z)} + \sqrt{1 + r(z)}\right)
\label{tz}
 \end{equation}
 Here $r(z)$ represents the relative contributions of dark energy and 
matter to the total energy density of the universe at redshift $z$,
 \begin{equation}
r(z) = \frac{\Omega_\Lambda}{\Omega_m} \frac{1}{(1 + z)^3}.
 \end{equation}
 For neutrinos with energy $E$ today, we can infer that they were 
generated from Majorons that decayed at redshift $z_E$, where $z_E$ is
given by 
 \bea
1+z_E = \frac{p_{\rm max}}{\sqrt{E^2-m_\nu^2}} ~~\textrm{with}~~p_{\rm max}\equiv \sqrt{\frac{m_J^2}{4}-m_\nu^2}
\label{omegaz}.
 \eea
 Using Eqs.~(\ref{tz}) and (\ref{omegaz}) to eliminate $t$ in favor of 
$E$ in Eq.~(\ref{nu0}), we can obtain an expression for the energy 
distribution of the extragalactic neutrinos from Majoron decay (see for 
example \cite{Cui:2017ytb}),
 \begin{equation} 
 \frac{d n^{\rm e.g.}_\nu}{d E} =  \mathcal{F} \Omega_{\rm dm} \frac{\rho_c}{m_J} 
 \frac{1}{\tau_J H_0 \sqrt{\Omega_\Lambda+\Omega_m (1+z_E)^3}}
e^{-\frac{t(z_E)}{\tau_J}} \frac{E}{E^2-m_\nu^2}~~(E\in [m_\nu, \frac{m_J}{2}]).
\label{nu1}
 \end{equation}
 
\subsection{Galactic Neutrinos }

If the Majoron lifetime is comparable to the age of universe, Majorons 
will clump inside the Milky Way galaxy. If the Majorons are sufficiently 
cold, their distribution in the galaxy will follow the cold dark matter 
profile. Since the time to traverse the galaxy is small compared to the 
age of the universe, any neutrinos we observe from Majoron decay in the 
galaxy must have arisen from decays in the recent past. The spectrum of 
the galactic neutrinos produced from Majoron decays is therefore 
monochromatic, with energy $m_J/2$. The flux of these neutrinos in the 
neighborhood of the earth is given by the line of sight integral,
 \bea
 n^{\rm gal}_{\nu} c =  
\mathcal{F} \frac{ r_\odot }{\tau_J}\frac{\rho_\odot}{m_J} \bar{J}
e^{-{t_0}/{\tau_J}},
 \eea
 where $r_\odot=8.33\,\rm kpc$ is the distance from the earth to the 
galactic center and $\rho_\odot=0.3\,\textrm{GeV/cm}^3$ is the dark 
matter density at the position of the Earth. Here the dimensionless 
factor $\bar J$ corresponds to the $J$ factor for the NFW 
profile~\cite{Navarro:1996gj} averaged over the Milky Way galaxy,
 \bea
\bar{J} =\frac{1}{4\pi}\int_{-\pi/2}^{\pi/2} db \cos b \int_0^{2\pi} 
d\ell J^{\rm NFW}(b,\ell)
\approx 2.19 \; .
 \eea
 Here we have used galactic coordinates $(d,b,\ell)$ with distance from 
the earth $d$, galactic latitude $b$ and longitude $\ell$. In this coordinate, 
we can express any vector in the $(x, y ,z)$ basis as:
\bea
x=d \cos b \cos \ell ~,~y=d \cos b \sin \ell~,~z=d \sin b.
\eea
The earth 
lies at the center of this coordinate system while the galactic center 
is at $x=r_\odot$ and $y=z=0$. This means the whole galactic plane corresponds to $z=0$ plane.  $J^{\rm NFW}(b,\ell)$ is the $J$ factor for the 
NFW profile expressed as a function of $b $ and $\ell$ 
\cite{Cirelli:2010xx}.

\section{Direct Detection in Neutrino Capture Experiments 
\label{s.signal}}

 The neutrinos from Majoron decay can be directly detected in 
experiments based on neutrino capture on nuclei. In this section we 
determine the size of the signal in experiments based on neutrino 
capture on tritium, with a focus on the PTOLEMY experiment.

 The PTOLEMY experiment has been designed to detect the C$\nu$B. The 
target consists of 100 grams of tritium coated on a graphene substrate. 
Tritium undergoes beta decay to He$^3$ with a lifetime of 12.32 years, 
liberating 18.6 keV of energy. This decay can also be induced by 
incident electron neutrinos, in which case the liberated electrons will 
have an energy that exceeds the end point energy of the beta decay 
process, allowing them to be distinguished. In the case of a 
non-relativistic neutrino with mass $m_\nu$, the energy of the liberated 
electrons will exceed the end point energy by $2 m_\nu$. Since PTOLEMY 
has an energy resolution of about $0.15$ eV, and the C$\nu$B neutrinos 
are nonrelativistic, it is sensitive to C$\nu$B neutrinos with masses 
above $0.1$ eV. In contrast, the neutrinos from the decays of Majorons 
with masses in the eV range and lifetimes of order the age of the 
universe are expected to be relativistic. We therefore expect that 
PTOLEMY will be able to detect these neutrinos, even if their masses lie 
well below $0.1$ eV, provided their energies today are above $0.2$ eV.
 
For low energy 
neutrinos the cross section $\sigma$ scales as the inverse of the 
relative velocity $v$, so that the product $\sigma v$ is a constant. The 
unpolarized capture rate on tritium for electron neutrinos is 
given by \cite{Long:2014zva}
\bea\label{eq:XS}
n_\nu\sigma v_\nu &=& \left[ (1- v_{\nu}) n_{\nu_{h_R}} + (1+v_{\nu}) n_{\nu_{h_L}}\right] \bar\sigma,\nn
\bar\sigma  &=& 3.83 \times 10^{-45} {\rm cm^2} \;,
\eea
 where $v_\nu$, the magnitude of the neutrino velocity, is close to 1 
for relativistic neutrinos. The number density of left(right)-helical 
neutrinos is denoted as $n_{\nu_{h_L(h_R)}}$.  If neutrinos are Majorana 
(Dirac), the number density of neutrinos from Majoron decay is 
$n_{\nu_{h_L}}=n_{\nu_{h_R}}= n^{J}_\nu$ 
($n_{\nu_{h_L}}=\frac{1}{2}n^J_\nu$\,,\,$n_{\nu_{h_R}}=\frac{1}{2}n^J_\nu$).
 Eq.~(\ref{eq:XS}) is valid for incident neutrinos with energies below a 
few keV. It is clear from this formula that, for a given Majoron mass 
and lifetime, the signal rate from Majorana neutrinos (denoted as 
$R_{\rm M}$) is larger by a factor of two than the rate from Dirac 
neutrinos (denoted as $R_{\rm D}$), $R_{\rm M}= 2R_{D}$. Neutrinos from 
galactic and extragalactic Majoron decays exhibit characteristic energy 
distributions, and the signal event rate at PTOLEMY will maintain the 
relation $R_{\rm M}= 2R_{D}$ throughout the entire spectrum.

 Only neutrinos in the electron flavor can be captured on tritium. Since 
the neutrinos produced in Majoron decay are in mass eigenstates, the 
event rate in PTOLEMY depends on the neutrino masses and mixing angles. 
The probability that a neutrino from Majoron decay is in the eigenstate 
with mass $m_i$ is given by,
 \begin{equation}
P_i = \frac{m_i^2}{m_1^2 + m_2^2 + m_3^2}  \;.
 \end{equation}  
 The capture cross section for a neutrino in the $i$'th mass eigenstate 
is smaller by a factor of $|U_{ei}|^2$ than that for an electron 
neutrino. It follows that the overall number of events from Majoron 
decay is suppressed by an overall factor of $\sum_i |U_{ei}^2| P_i$ 
relative to that from an equivalent flux of purely electron neutrinos. 
The current best fits to (three flavor) neutrino data yield $|U_{e1}|^2 
= 0.68$, $|U_{e2}|^2 = 0.30$ and $|U_{e3}|^2 = 0.022$. The solar and 
atmospheric mass splittings are given by $0.0086$ eV and $0.050$ eV 
respectively \cite{Esteban:2016qun}.\footnote{We use the latest 2018 data 
for normal hierarchy from NuFIT 3.2 (2018), www.nu-fit.org.} Then, if 
the lightest neutrino is massless, the suppression factor is about 
$0.03$ in the case of a normal hierarchy, and about $1/2$ in the case of 
an inverted hierarchy. If neutrinos are quasi-degenerate, the 
suppression factor is about $1/3$.
 
\begin{figure}[!t]
\includegraphics[width=0.48\linewidth]{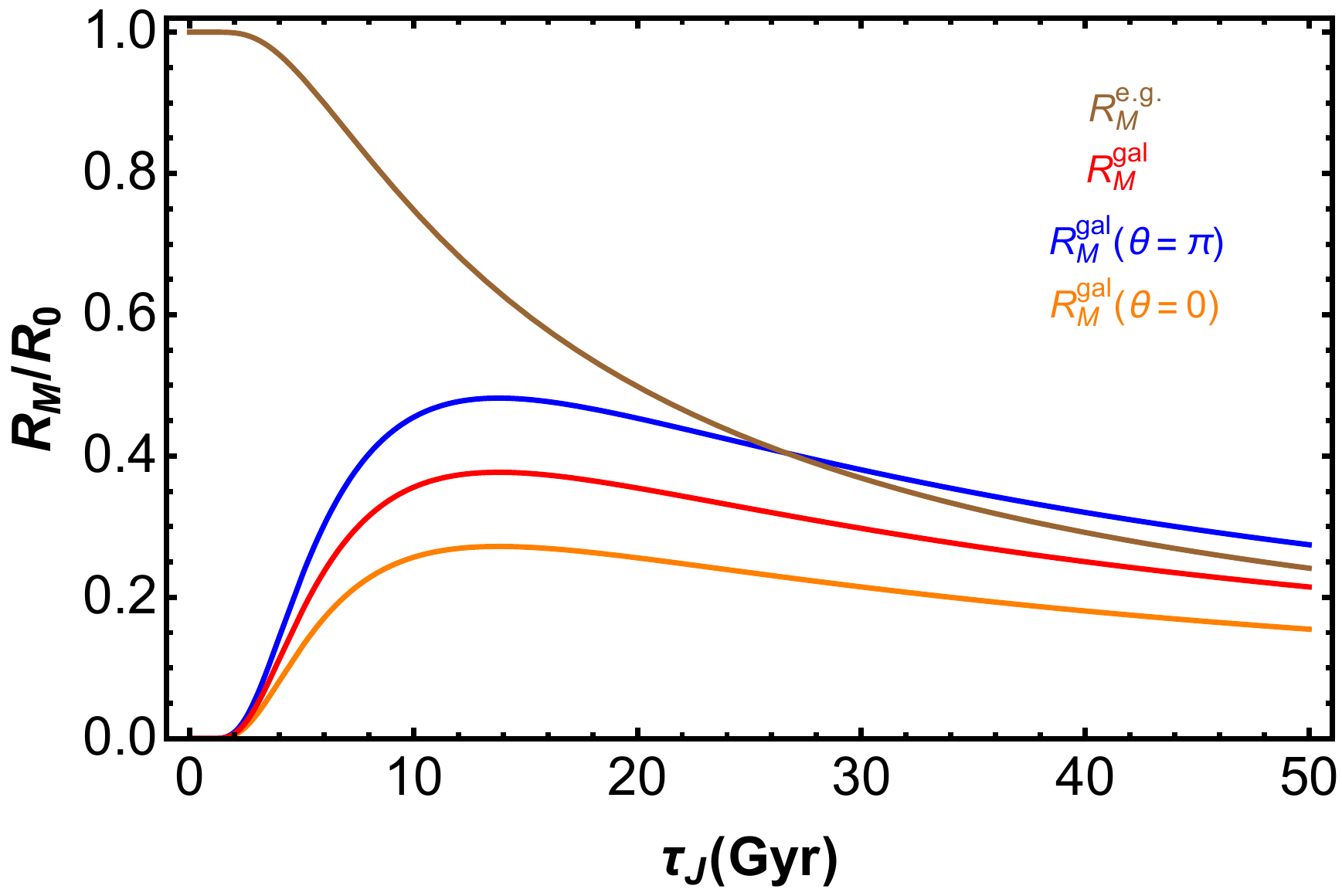}
\caption{ The signal rate normalized to $R_0$ [eq.~(\ref{eq:iso_result})] as a function of Majoron life time ($\tau_J$) from the extragalactic neutrino source ($R_M^{\rm e.g.}$), the averaged galactic neutrinos ($R_M^{\rm gal}$), galactic neutrinos with tritium spin polarized towards [$R_M^{\rm gal}(\theta=0)$] and away [$R_M^{\rm gal}(\theta=\pi)$] from the GC. }
\label{fig:signal}
\end{figure}

 The number of extragalactic (galactic) neutrino events per year expected in 
PTOLEMY, in the case of Majorana neutrinos, is given by (see Fig.~\ref{fig:signal})
 \begin{equation}\label{eq:event_rate}
R^{\rm e.g.}_M = R_0 \left( 1 - e^{-t_0/\tau_J}\right)~~,~~R^{\rm gal}_M = R_0 \frac{r_\odot 
\rho_\odot}{\tau_J\Omega_{\rm dm}\rho_c}\bar Je^{- t_0/\tau_J} 
 \end{equation}
 where $R_0$ is given by
 \bea\label{eq:iso_result}
 R_0 &=& N_{H} 2n^J_{\nu0}\sum_i |U_{ei}^2| P_i\bar \sigma \nn
     &=& \frac{3.1}{{\rm year}} \,
  \frac{\mathcal{F}}{5\%} \frac{{\rm eV}}{m_J} 
 \frac{\sum_i |U_{ei}^2| P_i}{1/3} 
\frac{M_T}{{\rm 100\, g}}
\frac{\bar\sigma }{ 3.83 \times 10^{-45} {\rm cm^2}}.
 \eea
 Here $N_{\rm H}$ is the number of tritium nuclei in the detector and 
$M_T$ is the mass of the tritium target. We see that in the case of 
quasi-degenerate and inverted spectra, we can expect to see as many as a 
few events per year, provided the neutrino energies are above the 
detector threshold. This falls to one event every few years in the case 
of a normal hierarchy. Note that this signal rate is comparable to the 
expected event rate for the standard C$\nu$B, which corresponds to 4(8) 
events at PTOLEMY if neutrinos are Dirac 
(Majorana).\footnote{Gravitational clustering of nonrelativistic 
neutrinos is expected to modify the C$\nu$B event rate, but only by 
order one factors \cite{Ringwald:2004np, Zhang:2017ljh}.}

\subsection{Extragalactic neutrino spectrum}

Since extragalactic neutrinos are approximately isotropic, we can use 
the unpolarized cross-section in Eq.~(\ref{eq:XS}). By combining 
Eq.~(\ref{nu1}) and Eq.~(\ref{eq:XS}), we obtain the event rate for 
Majorana neutrinos,
 \bea\label{eq:eg_signal}
\frac{1}{R_0} \frac{d R^{\rm e.g.}_M}{dE} =  \frac{1}{\tau_J H_0 \sqrt{\Omega_\Lambda+\Omega_m (1+z_E)^3}}
e^{-\frac{ t(z_E)}{\tau_J}} \frac{E}{E^2-m_\nu^2}~~(E\in [m_\nu, \frac{m_J}{2}]),
 \eea
 where $z_E$ is defined in Eq.~(\ref{omegaz}).

\begin{figure}[!t]
\includegraphics[width=0.48\linewidth]{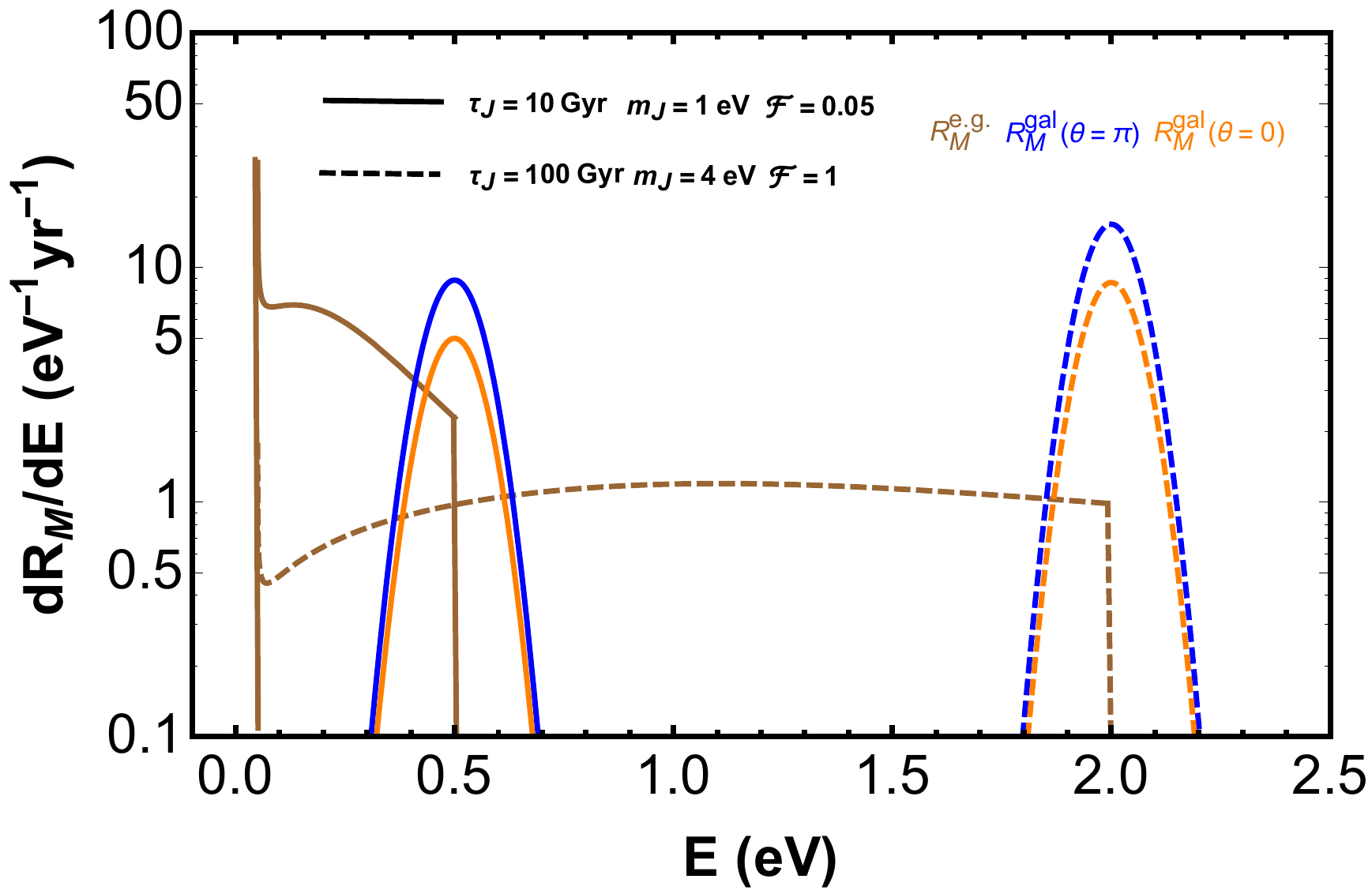}
\caption{  
 This plot shows $dR_M/dE$ as a function of the neutrino energy $E$ for 
two benchmark points, both with $m_\nu=0.05\,$eV and 100 g tritium target. Solid lines represent 
the benchmark ($m_J=1\,$eV, $\tau_J=10\,$Gyr and $\mathcal{F} =0.05$) 
while the dashed lines show the benchmark ($m_J=4\,$eV, 
$\tau_J=100\,$Gyr and $\mathcal{F} =1$). We show the energy spectra from 
extragalactic neutrino decays (brown), galactic neutrino decays with 
tritium spin polarized towards ($\theta=0$, orange) and away 
($\theta=\pi$, blue) from the galactic center. Here we apply 
$\sigma=0.064$ eV (from the experimental resolution $\Delta=\sqrt{8\ln2} 
\sigma=0.15$ eV) Gaussian smearing of galactic neutrinos. These results 
are for Majorana neutrinos. The rates for Dirac neutrinos are half of 
those in the Majorana case. }
 \label{fig:spectrum}
\end{figure}

 The spectrum of neutrinos from extragalactic Majoron decay is shown in 
Fig.~\ref{fig:spectrum}. For a given Majoron 
lifetime, the spectrum has sharp edges at $E={m_J}/{2}$ and $E=m_\nu$, 
with a continuous distribution connecting these two points. The peak at 
$E=m_\nu$ corresponds to the fact that neutrinos from early decays of 
Majorons are all nonrelativistic with energies close to the 
neutrino mass.

By integrating Eq.~(\ref{eq:eg_signal}), $R^{\rm e.g.}_M = \int dE 
\frac{d R^{\rm e.g.}_M}{ dE} = R_0 (1-e^{-t_0/\tau_J})$, we reproduce 
the result for the total event rate.  If the lifetime is smaller than 
the age of universe ($\tau_J < t_0$), most of the Majorons will have 
decayed. Then the event rate is expected to be close to $R_0$, $R^{\rm 
e.g.}_M \approx R_0$, provided the neutrinos retain enough energy to 
pass the detector threshold. If, however, the lifetime is longer than 
the age of the universe, the fraction of Majorons that have decayed 
before today is smaller. Then the ratio $R^{\rm e.g.}_M/R_0$ gets smaller 
as $\tau_J$ increases, as can be seen in Fig.~\ref{fig:signal}.

\subsection{Polarized tritium and the galactic neutrino spectrum}

The capture cross section of neutrinos on tritium depends on the 
relative angle between the incoming neutrino velocity and the spin of 
the tritium nucleus. As discussed in \cite{Lisanti:2014pqa}, this means 
that if the tritium used in PTOLEMY is polarized, we can obtain 
information about the direction of the neutrino signal. This in turn can 
be used to obtain information about the distribution of the Majoron 
component of dark matter in the galaxy. Since the galactic neutrinos 
will have energies of order an eV, they can be treated as 
ultrarelativistic. Averaging over the angles of the outgoing electrons, 
one obtains the polarized capture rate for relativistic electron neutrinos 
as~\cite{Lisanti:2014pqa}
 \bea\label{eq:XS_polarized}
n_\nu\sigma({\mathbf{\hat s}}, {\mathbf{\hat v}})v= 2 n_{\nu_{h_L}}  \bar \sigma(1+B\, {\mathbf{\hat s}}\cdot {\mathbf{\hat v}}) \;.
 \eea
 Here the parameter $B\approx 0.99$ incorporates the relevant couplings 
and form factors. The unit vector in the direction of the tritium spin is 
denoted as $\mathbf{\hat s}$, while the direction of the incoming 
neutrino is $\mathbf{\hat v}$. As mentioned earlier,  
$n_{\nu_{h_L}}=n^J_\nu$ if neutrinos are Majorana, while 
$n_{\nu_{h_L}}=\frac{1}{2}n^J_\nu$ if neutrinos are Dirac.
 Since the extragalactic source (discussed in section 
\ref{sec:e.g._neutrino}) is isotropic, the result for that case simply 
reduces to Eq.~(\ref{eq:XS}) after averaging the tritium spin in the 
relativistic limit. However, because the galactic distribution of 
Majorons follows an NFW profile, interesting directional information is 
encoded in term proportional to ${\mathbf{\hat s}}\cdot {\mathbf{\hat 
v}}$ in Eq.~(\ref{eq:XS_polarized}).

We wish to determine the dependence of the signal on the orientation of 
the spin of the tritium nuclei relative to the galactic center. We begin 
by writing ${\mathbf{\hat s}}\cdot {\mathbf{\hat v}}$ in terms of 
galactic coordinates,
 \bea
 \label{eq:general_dot_prod}
\mathbf{\hat s}\cdot \mathbf{\hat v} = 
\cos b_{\mathbf{\hat s}} \cos b_{\mathbf{\hat v}}\cos( \ell_{\mathbf{\hat v}}-\ell_{\mathbf{\hat s}} )+ \sin b_{\mathbf{\hat s}} \sin b_{\mathbf{\hat v}} ,
 \eea
 where ${\mathbf{\hat s}}({\mathbf{\hat v}})=\cos b_{{\mathbf{\hat 
s}}({\mathbf{\hat v}})}\cos \ell_{{\mathbf{\hat s}}({\mathbf{\hat v}})}{\mathbf{\hat e}_x}+\cos b_{{\mathbf{\hat 
s}}({\mathbf{\hat v}})}\sin \ell_{{\mathbf{\hat s}}({\mathbf{\hat v}})}{\mathbf{\hat e}_y}+\sin b_{{\mathbf{\hat 
s}}({\mathbf{\hat v}})}{\mathbf{\hat e}_z}$ with ${\mathbf{\hat e}_{x,y,z}}$ being the unit vector along $x,y,z$ direction. We also define $\theta$ as the angle between the 
tritium spin and the unit vector pointing towards the galactic center 
(${\mathbf{\hat e}_x}$),
 \bea\label{eq:cos_theta_0}
\cos \theta\equiv\mathbf{\hat s}\cdot {\mathbf{\hat e}_x} =\cos b_{\mathbf{\hat s}}\cos \ell_{\mathbf{\hat s}} .
 \eea
 Since the form of the NFW distribution is invariant under rotations 
along the line connecting the earth and the galactic center, the final 
result should also be unchanged under such a rotation. This allows us to 
choose coordinates such that $\ell_{\mathbf{\hat s}}=0 
~\textrm{or}~\pi$. Combining Eqs.~(\ref{eq:general_dot_prod}) and 
(\ref{eq:cos_theta_0}), we obtain
 \bea
\mathbf{\hat s}\cdot \mathbf{\hat v} &= & \left\{
\begin{array}{ll}
\cos b_{\mathbf{\hat s}} \cos b_{\mathbf{\hat v}}\cos \ell_{\mathbf{\hat v}} + \sin b_{\mathbf{\hat s}} \sin b_{\mathbf{\hat v}} ~~~~~(\ell_{\mathbf{\hat s}}=0) \no\\
-\cos b_{\mathbf{\hat s}} \cos b_{\mathbf{\hat v}}\cos \ell_{\mathbf{\hat v}} +\sin b_{\mathbf{\hat s}} \sin b_{\mathbf{\hat v}}~~~(\ell_{\mathbf{\hat s}}=\pi)
\end{array}
\right. \\
&=&\cos \theta \cos b_{\mathbf{\hat v}}\cos \ell_{\mathbf{\hat v}} + \sin  \theta  \sin b_{\mathbf{\hat v}}.
 \eea

 In this case the rate when the spin is oriented at an angle $\theta$ with Majorana neutrinos
is given by
 \bea\label{eq: gal_rate_angle}
\frac{R_M^{\rm gal}(\theta)}{R_0}&=&\frac{r_\odot 
\rho_\odot}{\tau_J\Omega_{\rm dm}\rho_c}e^{- t_0/\tau_J} \left[\bar J+\frac{B}{4\pi}\int_{-\pi/2}^{\pi/2} db  \int_0^{2\pi} d\ell \cos b(-\cos \theta\cos b  \cos \ell - \sin \theta\sin b ) J^{\rm NFW} (b,\ell)\right]\nonumber\\
&\approx&\frac{r_\odot 
\rho_\odot}{\tau_J\Omega_{\rm dm}\rho_c}e^{- t_0/\tau_J}   (2.19- 0.60 \cos \theta ).
\eea

 It is clear from Eq.~(\ref{eq: gal_rate_angle}) that the galactic 
signal varies by almost a factor of two depending on whether the spin of 
the polarized tritium points towards or away from the galactic center,
 \bea
\frac{R_M^{\rm gal}(\theta=\pi)}{R_M^{\rm gal}(\theta=0)}\approx 1.8.
 \eea
 This is a characteristic feature of the galactic neutrino signal. 
PTOLEMY with polarized tritium can test this prediction by measuring the 
modulation associated with the orientation of the 
spin.\footnote{Polarized tritium can also be used to measure the annual 
modulation of PTOLEMY signals, both in the context of the standard 
C$\nu$B \cite{Safdi:2014rza}, and of new physics \cite{Huang:2016qmh}.}

In principle, all galactic neutrinos have the same energy $E=m_J/2$. 
However, due to experimental resolution, the actual spectrum of galactic 
neutrinos will be smeared out. To obtain a realistic spectrum of 
galactic neutrinos, we apply a Gaussian smearing with the full width at 
half maximum (FWHM) $\Delta$ defined as
 \bea
\Delta=\sqrt{8\ln 2} \sigma,
 \eea
 where $\sigma$ is the standard deviation of the Gaussian. The expected 
energy resolution of PTOLEMY, $\Delta=0.15$ eV, corresponds to a 
standard deviation $\sigma=0.064$ eV. With this Gaussian smearing, the 
galactic neutrinos spectrum is given by
 \bea
\frac{1}{R_0} \frac{d R_M^{\rm gal}(\theta)}{d E}
 =
\frac{1}{\sqrt{2\pi}\sigma}\frac{n^{\rm gal}_0}{n^J_{\nu0}}\left(2.19-0.60\cos \theta\right)\textrm{exp}[-\frac{(E-\frac{m_J}{2})^2}{2\sigma^2}],
 \eea
 where 
 \begin{equation}
\frac{n^{\rm gal}_0}{n^J_{\nu0}}=\frac{r_\odot 
\rho_\odot}{\tau_J\Omega_{\rm dm}\rho_c}e^{- t_0/\tau_J} \;.
 \end{equation} 
 The galactic neutrino spectrum is shown in Fig.~\ref{fig:spectrum}. 
From the above expression, we see that the total signal rate from 
galactic neutrinos is proportional to $e^{-t_0/\tau_J}/\tau_J$. 
Therefore, $R_M^{\rm gal}$ is peaked at $\tau_J=t_0$, as can be seen in 
Fig.~\ref{fig:signal}). If $\tau_J\ll t_0$ the event rate is small due 
to the galactic neutrino density being suppressed by $e^{-t_0/\tau_J}$. 
On the other hand, if $\tau_J \gg t_0$, the signal is suppressed because 
of the small decay rate.

Comparing the signals from extragalactic and galactic Majoron decay, we 
see that both signals can be easily distinguished from each other, and 
from C$\nu$B signals. C$\nu$B signals accumulate close to $E=m_\nu$ 
while the energy distribution of neutrinos from extragalactic decays 
exhibits a distinctive spectrum, as shown in Fig.~\ref{fig:spectrum}. 
Galactic signals can be distinguished by their monochromatic nature, 
and, if the tritium target is polarized, by employing directional 
information.

The directional dependence of the signal also encodes information about 
the dark matter distribution in the Milky Way galaxy. Once $R^{\rm 
gal}_M(\theta)$ is measured, the ratio of the $\theta$ independent term 
to the coefficient of the $\cos\theta$ term captures the ratio of the 
average of the dark matter profile to the weighted average along the 
line sight [see Eq.~(\ref{eq: gal_rate_angle})]. Measuring this ratio 
can help discriminate between models of the dark matter profile in the 
Milky Way.

 \subsection{Benchmark points}
 
For concreteness we consider two benchmark points. We first choose 
$m_J=1\,$eV and a lifetime that is lightly shorter than the age of the 
universe, $\tau_J=10\,$Gyr. In this case the current bounds from 
decaying dark matter indicate that Majorons can only contribute about 
$5\%$ of the total dark matter abundance \cite{Poulin:2016nat}. 
Therefore, the first benchmark point we choose is BP1: $m_J=1\,$eV, 
$\tau_J=10\,$Gyr and $\mathcal F =0.05$. From the analysis in 
\cite{Poulin:2016nat}, if the lifetime of the Majoron is sufficiently 
longer than the age of the universe, say $\tau_J=100\,$Gyr, it can 
constitute all of the dark matter abundance, so that $\mathcal F=1$. 
Therefore we consider another benchmark BP2: $m_J=4\,$eV, 
$\tau_J=100\,$Gyr and $\mathcal F =1$. For these two benchmark points 
with quasi-degenerate neutrino masses, we obtain the total signal rates 
from extragalactic decays [Eq.~(\ref{eq:event_rate})] and galactic 
decays with a polarized tritium target [Eq.~(\ref{eq: gal_rate_angle})] 
as
 \bea
\textrm{BP1}:&~&R_M^{\rm e.g.}=2.3 /(100 \textrm{g}\cdot \textrm{yr}) ~~,~~R_M^{\rm gal}(\theta=\pi)=1.4 /(100 \textrm{g}\cdot \textrm{yr}) \nonumber\\
\textrm{BP2}:&~&R_M^{\rm e.g.}=2.0 /(100 \textrm{g}\cdot \textrm{yr})  ~~,~~R_M^{\rm gal}(\theta=\pi)=2.4 /(100 \textrm{g}\cdot \textrm{yr}) .
 \eea

The dominant background at PTOLEMY arises from regular beta decay events 
in which the electron carries away almost all the liberated energy. We 
expect that this background falls off very quickly for $E>0.2\,$eV in 
Fig.~\ref{fig:spectrum}. The basis of this expectation is that PTOLEMY 
experiments have sensitivity to C$\nu$B neutrinos with $m_\nu\gtrsim 
0.1\,$eV, which means that the experiment must be able to discriminate 
events with electron energy $0.2\,$eV or more above the end point of the 
beta decay spectrum. Following this assumption, we obtain the sum of 
extragalactic and galactic event rates above $E\gtrsim 0.2\,$eV to be 
$R_M^{\rm tot}=2.7$ for BP1 and $R_M^{\rm tot}=4.3$ for BP2 with 100 
gram tritium and one year exposure.  Therefore, we expect to have a few 
signal events with a specific energy spectrum in the small background 
range with just one year of data. Clearly, observing such a striking 
signal would provide strong evidence of Majoron decays.

\section{Conclusion\label{s.conclusion}}

In this paper we have analyzed the prospects for discovery of a 
secondary cosmic neutrino background originating from Majoron decays at 
future detectors based on neutrino capture on tritium, such as PTOLEMY. 
We have studied the bounds on this scenario and determined the event 
rate and energy distribution of the signal as a function of the Majoron 
mass and lifetime. We find that for Majoron masses at the eV scale and 
lifetimes of order the age of the universe, these detectors will be 
sensitive to this class of models with 100 gram-years of data. We have 
considered two distinct sources of this signal - galactic and 
extragalactic, and find that each has a distinctive spectrum that can be 
distinguished using the data. Finally, we have explored the possibility 
of using polarized tritium to obtain directional information about the 
incoming neutrinos. This can be used to discriminate between neutrinos 
with galactic and extragalactic origins, and thereby shed light on the 
distribution of dark matter in our galaxy.
\\

{\bf Note added:} While this paper was being completed, we received 
Ref.~\cite{McKeen:2018xyz}, which discusses ideas similar to those 
considered here.

\acknowledgments
 We would like to thank Oren Slone for useful discussions. ZC, PD and MG 
are supported in part by the National Science Foundation under Grant 
Number PHY-1620074. ZC would like to thank the Fermilab Theory Group for 
hospitality during the completion of this work. ZC's stay at Fermilab 
was supported by the Fermilab Intensity Frontier Fellowship and by the 
Visiting Scholars Award \#17-S-02 from the Universities Research 
Association.

\bibliographystyle{utphys}
 
\bibliography{CnuBbib}

\end{document}